# High Intrinsic Catalytic Activity of Two-Dimensional Boron Monolayers for Hydrogen Evolution Reaction

Li Shi[a#], Chongyi Ling[a#], Yixin Ouyang[a] and Jinlan Wang[*a,b]

[a] Department of Physics, Southeast University, Nanjing 211189, China
[b] Synergetic Innovation Center for Quantum Effects and Applications (SICQEA), Hunan Normal University, Changsha 410081, China
#These authors contributed equally to this work
*Corresponding Author: jlwang@seu.edu.cn

**Abstract:** Two-dimensional (2D) boron monolayers have been successfully synthesized on silver substrate very recently. Their potential application is thus of great significance. In this work, we explore the possibility of boron monolayers (BMs) as electrocatalysts for hydrogen evolution reaction (HER) by first-principle method. Our calculations show that the BMs are active catalysts for HER with nearly zero free energy of hydrogen adsorption, metallic conductivity and plenty of active sites in the basal plane. The effect of the substrate on the HER activity is further assessed. It is found that the substrate has a positive effect on the HER performance caused by the competitive effect of mismatch strain and charge transfer. The indepth understanding of the structure dependent HER activity is also provided.

**Introduction**

With the boom in graphene research, ultrathin two-dimensional (2D) materials are attracting dramatically increasing interest over the past decade.[1-7] Plenty of 2D materials have been prepared beyond graphene, such as hexagonal boron nitride (*h*-BN),[8] transition-metal dichalcogenides (TMDs),[5, 9, 10] graphitic carbon nitride (g-$C_3N_4$),[11] silicene,[12] metal organic frameworks (MOFs),[13, 14] black phosphorus (BP),[15, 16] early transition-metal carbides (MXenes)[17-19] and so on. Owing to their unique structure and extraordinary properties, 2D materials have become a key class of materials in chemistry, physics as well as materials science, and more importantly, exhibited potential applications in various fields, including electronics,[20] supercapacitors,[21] batteries,[22] catalysis,[5] sensors[23]. Among these 2D materials, single-element 2D nanosheets are still very rare.

Boron (B) is the fifth element in the periodic table and the bonding formed between B atoms is more complicated than that in carbon. A series of boron monolayers (BMs) with different structures have been proposed by theoretical predictions.[24, 25] Very recently, two parallel experimental works reported the synthesis of 3 kinds of atomically thin BMs, $\beta_{12}$, $\chi_3$ and trigonal type.[26, 27] Under this condition, the potential application of BMs is becoming extremely urgent and significant. Like other 2D materials, BMs have large surface areas, which provides enough sites for the adsorption of reactants and more importantly, most BMs are found to be metallic[25, 27], indicating that BMs may be good electrocatalysts for hydrogen evolution reaction (HER).

In this work, we select 5 kinds of BMs, $\beta_{12}$, $\chi_3$ and trigonal type[26, 27] which have already been synthesized experimentally as well as $\alpha_1$ and $\beta_1$[25], two most stable structures from computational prediction, to investigate their catalytic activity for HER. Our first-principle calculations show that the BMs are highly active catalysts for HER with the free energy of hydrogen adsorption ($\Delta G_H$) close to 0 eV. We further explore the influence of the substrate on the HER performance, as currently synthesized BMs are all grown on certain substrates experimentally. It is found that the HER performance of BM@Ag(111) will be improved due to the competitive effect of mismatch strain and charge transfer.

**Results and Discussion**

We first study the HER performance of the $\beta_{12}$, $\chi_3$ and trigonal BMs that have already been fabricated in laboratory. For $\beta_{12}$ BM, there are three possible sites for H adsorption, which are labeled as $\beta_{12}$-$S_1$, $\beta_{12}$-$S_2$ and $\beta_{12}$-$S_3$, respectively, as shown in Figure 1a. The calculated free energies of H adsorption at these sites are 0.10, 1.13 and 0.23 eV, respectively (Figure 1d). Thus, HER at $\beta_{12}$-$S_1$ site will present high activity, as the $|\Delta G_H|$ at this site is comparable to that of Pt (-0.09eV).[28] For $\chi_3$ BM, two distinct sites for H adsorption can be obtained, $\chi_3$-$S_1$ and $\chi_3$-$S_2$ (see Figure 1b). However, the $\Delta G_H$ for $\chi_3$-$S_2$ is 0.53 eV, indicating the weak binding strength of H at this site and making proton transfer difficult. Therefore, the $\chi_3$-$S_2$ site is not suitable for HER. Whereas, the $\Delta G_H$ for $\chi_3$-$S_1$ is only 0.02 eV, which is even much closer to 0 eV than that of Pt and $MoS_2$, two catalysts that have been proved to be with excellent catalytic activity by lots of experiment,[28-31] showing that the $\chi_3$-$S_1$ site owns ultra-high activity as the HER electrocatalyst. As for trigonal BM, its surface is simpler than the former two BMs due to the high symmetry and there is only 1 kind of site (see Figure 1c). The calculated $\Delta G_H$ is -0.46 eV, demonstrating that the strong binding strength of H would lead to the release of gaseous $H_2$ product slow, and therefore, the trigonal BM is not an ideal electrocatalyst for HER under relatively low hydrogen coverage (1/18 ML). It should be noted that the $\Delta G_H$ will increase with the increase of hydrogen coverage according to previous studies,[19, 28] so it is for BMs (Table S1). Therefore, the HER performance of trigonal BM will be improved under a higher hydrogen coverage. More specifically, In the range of 3/18 to 5/18 ML of hydrogen coverage, the trigonal BM presents excellent HER activity.

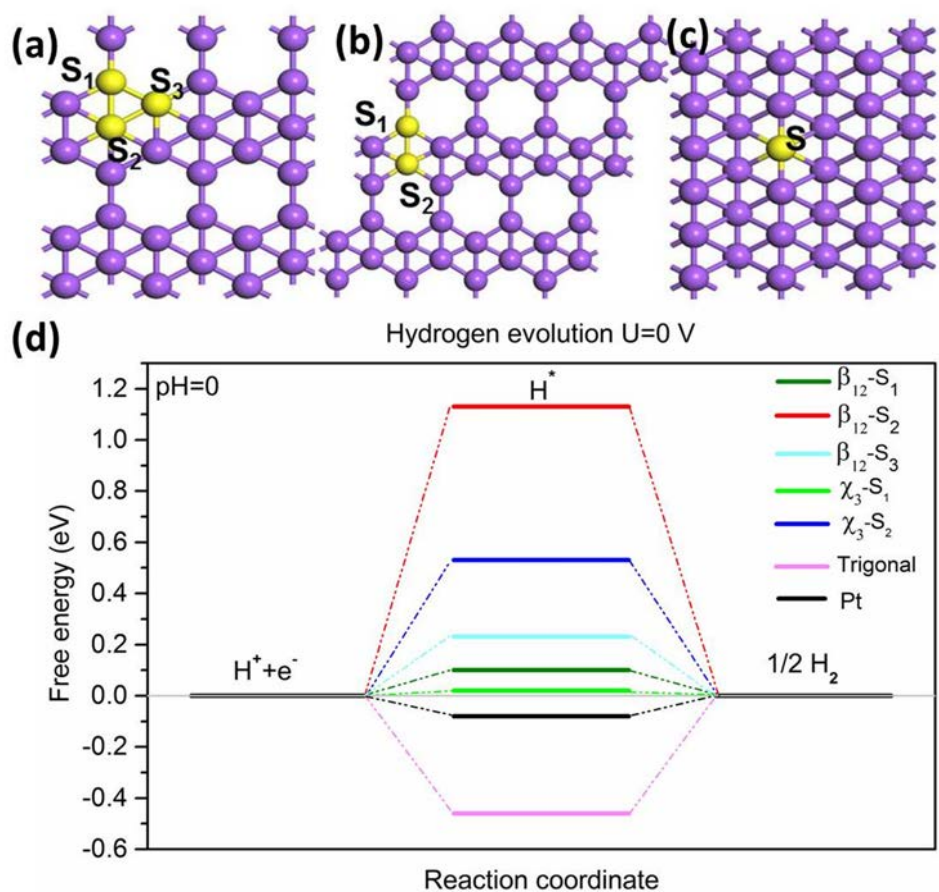

**Figure 1.** The structures of (a) $\beta_{12}$, (b) $\chi_3$ and (c) trigonal BM, respectively. The purple and yellow balls refer to the B atoms and disparate sites for H adsorption, respectively. (d) Calculated free energy diagram for hydrogen evolution of different active sites at a potential $U$ = 0 relative to the standard hydrogen electrode at pH=0.

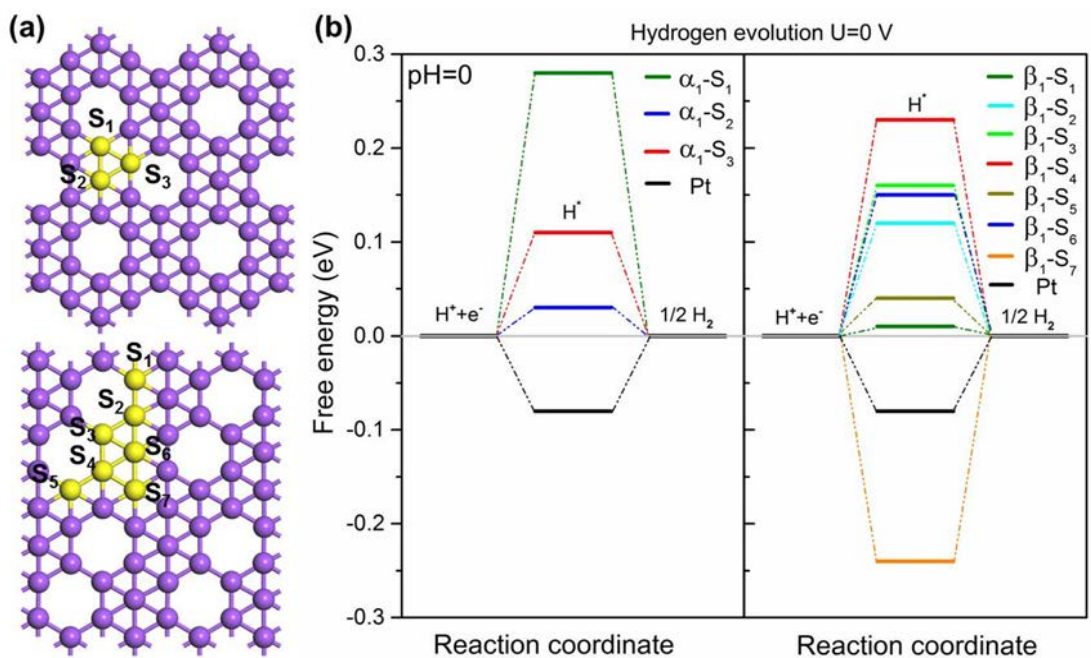

**Figure 2.** (a) The structures of $\alpha_1$ (top) and $\beta_1$ (bottom) BM. (b) The calculated free energy diagram for hydrogen evolution of different active sites at a potential $U$ = 0 relative to the standard hydrogen electrode at pH = 0. The purple and yellow balls refer to the B atoms and disparate sites for hydrogen adsorption, respectively.

Next we move on to an assessment of the HER performance of $\alpha_1$ and $\beta_1$ BMs which were predicted to be the most stable structures by first-principles calculations. As shown in Figure 2a, there are 3 and 7 different sites for H adsorption in $\alpha_1$ and $\beta_1$ BMs, respectively. The computed $\Delta G_H$ of these 10 sites are plotted in Figure 2b, where most of them own small $|\Delta G_H|$ comparable to or even smaller than that of Pt[28] except $\alpha_1$-$S_1$, $\beta_1$-$S_4$ and $\beta_1$-$S_7$. In particular, the $\Delta G_H$ for $\alpha_1$-$S_2$, $\beta_1$-$S_1$ and $\beta_1$-$S_5$ is just 0.03, 0.01 and 0.04 eV, respectively, indicative of the ultra-high activity for HER at these sites. Other sites, such as $\alpha_1$-$S_3$, $\beta_1$-$S_2$, $\beta_1$-$S_3$, and $\beta_1$-$S_6$, have comparable $|\Delta G_H|$ (0.11, 0.12, 0.16, and 0.15 eV, respectively) to that of Pt as well. Therefore, we can conclude that, $\alpha_1$

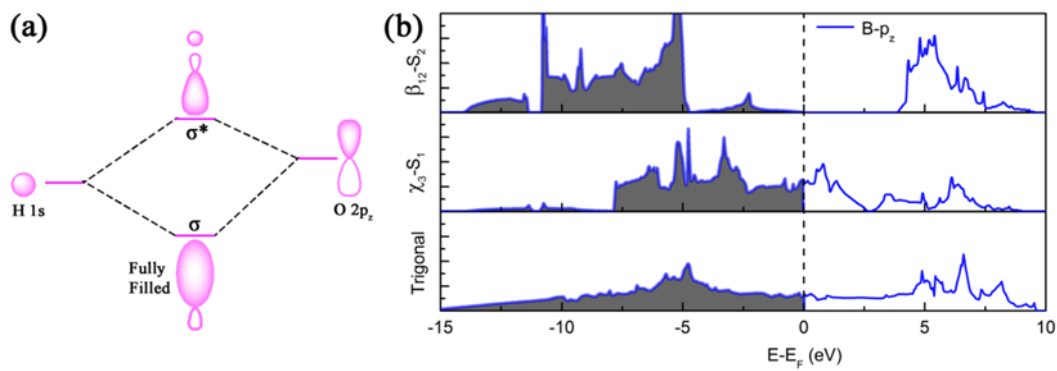

**Figure 3.** (a) The diagram of formation of H-B bonds on the surface of BM. (b) Projected $p_z$-orbital density of states for the B atom. The shaded area corresponds to filled states up to the Fermi level.

and $\beta_1$ BMs are both potential catalysts for HER.

Apart from catalytic activity, the stability is also vital to the catalysts. In a recent study, Tsai et al. used the degree of difficulty for $H_2X$ (X = S, Se) desorption to describe the HER stability of transition metal dichalcogenide ($MX_2$) by calculating the HX adsorption free energy.[32] In our cases, as the simplest boron hydride is $B_2H_6$, we use it as the reference system and calculate the free energy difference between the formation of $H_2$ and $B_2H_6$ on BMs ($\Delta G$) to determine which product is energetically more favorable. Herein, two most active sites, $\beta_1$-$S_1$ and $\beta_{12}$-$S_1$ with 4 and 3 possible desorption paths for $B_2H_6$ (Figure S1 in Supporting Information), respectively, are selected to study the HER stability of BMs. As shown in Table S2, the free energy differences are all rather negative (smaller than 1.60 eV), indicating that the adsorbed hydrogen atoms on boron monolayers much prefer to form $H_2$ rather than the $B_2H_6$. This can be easily understood that the release of $B_2H_6$ will lead to the formation of two B vacancies on BMs, which is unfavorable as boron is an electron-deficient element[24]. Therefore, boron monolayers are very stable when used as catalyst for HER.

Above results clearly demonstrate that different types of BMs have distinct HER performance. To understand the structure-dependent HER activity, we present a detailed analysis of the density of states combined with the molecular orbital theory. As mentioned in the section of computational details, the $\Delta G_H$ is determined by the H binding strength on the surface. For H adsorbs on the surface of BM, the H-B bonds are formed by the linear combination between the H 1s orbital and B $2p_z$ orbital. This combination results in a fully filled, low-energy bonding orbital (σ) and a partially filled, high-energy anti-bonding orbital (σ*) (Figure 3a). Moreover, the bonding strength can be described by bond order, which equals to half of the difference between the electron number of σ and σ* according to the molecule orbital theory. Thus, more electrons in B $2p_z$ orbital will lead to higher σ* occupancy and subsequently weaker binding strength of H-B bonds. Based on this principle, we plot the partial density of states (PDOS) of B $2p_z$ orbital at $\beta_{12}$-$S_2$, $\chi_3$-$S_1$ and the trigonal sites, which represents the weak, moderate and strong H-B bonding strength, respectively, as shown in Figure 3b. It is evidently observed that the occupancy of valence band gradually

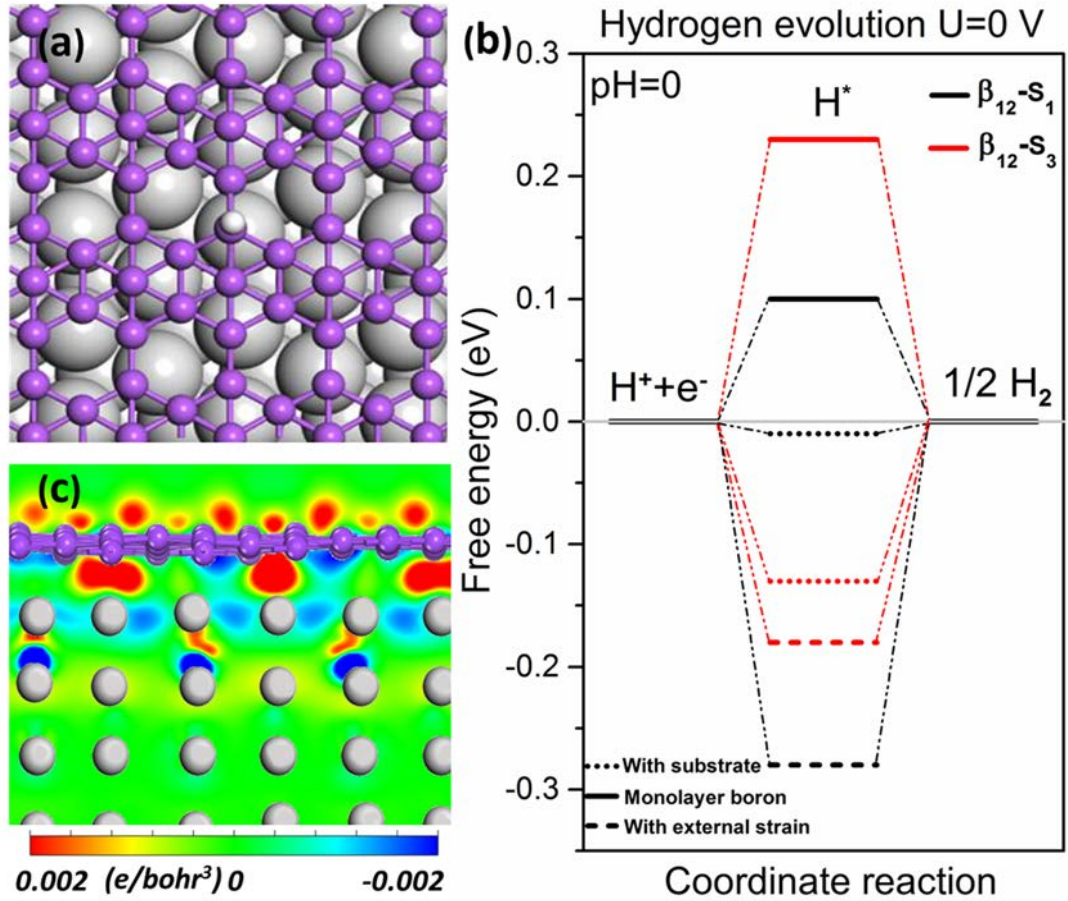

**Figure 4** (a) The structure of $\beta_{12}$ BM which is placed on Ag (111) surface. The purple and grey balls represent boron and silver atoms respectively. (b) Calculated free energy diagram for hydrogen evolution of $\beta_{12}$ BM with external strain and with Ag substrate at a potential $U$ = 0 V relative to the standard hydrogen electrode at pH = 0. The applied external strain is -1.2% along $a$ and +1.5% along $b$, exactly same to the values produced by the lattice mismatch between BM and Ag (111). (c) Differential charge density distributions of $\beta_{12}$ BM on Ag surface. Red and blue colors indicate the positive and negative values of electron.

decreases when the site changes from $\beta_{12}$-$S_2$ to $\chi_3$-$S_1$ to trigonal site. To gain more accurate analysis, we further compute the electron number of $2p_z$ orbital of these three sites by integrating the whole valence band (shadow area presented in the Figure 3b), where the corresponding electron numbers are 0.26, 0.21 and 0.19 e for $\beta_{12}$-$S_2$, $\alpha_3$-$S_1$ and the trigonal site, respectively. That is, the electron number of $2p_z$ orbital presents a gradually declining tendency from $\beta_{12}$-$S_2$ to $\chi_3$-$S_1$ to trigonal site. Therefore, the H-B bond at $\beta_{12}$-$S_2$ is expected to be the weakest, followed by that at $\chi_3$-$S_1$ and the trigonal site, which is in line with our computation above. In addition, the smallest free energies for all the BMs (except trigonal BM) are close to 0 eV, indicating that the H atom would not be trapped at a site.

Above conclusions drawn are all based on free-standing BMs. However, the fabricated BMs to date are all grown on certain substrate, so it is very necessary to uncover the influence of the substrate on the HER activity. Here, we select $\beta_{12}$ BM as the prototype (Figure 4a) considering the lattice matching and the computational cost. Moreover, as the $\beta_{12}$-$S_2$ site is not an active site for HER, only $\beta_{12}$-$S_1$ and $\beta_{12}$-$S_3$ sites are taken into consideration.

After relaxation, the structures of BM on the substrate are basically same as the isolated ones (Figure 4a and Figure 1a). However, the $\Delta G_H$ has an evident change, as shown in Figure 4b. A general tendency is observed that the $\Delta G_H$ is decreased for both sites of BM on substrate (-0.01 and -0.13 eV for $\beta_{12}$-$S_1$ and $\beta_{12}$-$S_3$ sites, respectively) as compared with that of isolated ones (0.10 and 0.23eV for $\beta_{12}$-$S_1$ and $\beta_{12}$-$S_3$ sites, respectively). Therefore, the BM@Ag(111) actually presents improved HER activity as compared with that of free-standing BM. The decrease of $\Delta G_H$ can be ascribed to two aspects: mismatch strain and charge transfer between $\beta_{12}$ BM and Ag substrate. First of all, owing to the mismatch between the lattice parameter of Ag (111) and BM, the BM on Ag(111) substrate is actually strained (-1.2% in *a* direction and +1.5% in *b* direction in our computational model). In fact, previous studies have shown that extra strain has significant effect on the catalytic performance of 2D materials.[19, 33, 34] In order to evaluate the influence of the strain, we calculate the $\Delta G_H$ of free-standing BM under a biaxial strain of -1.2% along *a* and +1.5% along *b*, which is exact the same as the mismatch strain in BM@Ag(111) computational model. As shown in Figure 4b, the $\Delta G_H$ of strained free-standing BM has a significant decrease, from 0.10 to -0.28 eV and 0.23 to -0.18 eV for $S_1$ and $S_3$ sites, respectively. Meanwhile, the weak interaction between the BM and Ag substrate causes charge transfer from Ag substrate to BM (Figure 4c) and the $S_1$ site gains 0.02 e more charge than the $S_3$ site. According to our previous study,[19] the extra charge will occupy the anti-bonding orbital formed by the combination of H 1s orbital and B 2pz orbital, leading to lower bond order and weaker hydrogen bonding strength (Figure S2). Therefore, $\Delta G_H$ of both sites on BM@Ag(111) will increase and the $S_1$ site has a larger change than $S_3$ site as compared with that of strained free-standing BM (-0.28 to -0.01 eV and -0.18 to -0.13 eV for $S_1$ and $S_3$ sites, respectively). The competitive effect of charge transfer and mismatch strain between BM and Ag substrate make the $\Delta G_H$ of both sites on BM@Ag(111) actually present improved HER performance than that of pure BM. Therefore, BM@Ag(111) can be expected as an ideal HER catalyst as well.

## Conclusion

In summary, we have studied the possibility of BMs as potential electrocatalysts for HER. Our calculations demonstrate that the BMs are highly active for HER with nearly zero $\Delta G_H$, metallic conductivity and plenty of active sites in the basal plane. The high HER activity of BMs is very robust even supported on silver substrate, suggesting BMs are ideal catalysts for HER. This study opens a new window for the application of 2D boron monolayers and provides a new class of candidates for metal-free catalyst for HER.

## Computational details

The first-principles calculations were performed by using projector-augmented wave (PAW)[35] pseudopotential in conjunction with the Perdew-Burke-Ernzerhof generalized gradient approximation (PBE-GGA)[36, 37], as implemented in Vienna ab initio simulation package.[38, 39] The effect of van der Waals (vdW) interactions was included for weak interaction. The energy cutoff for the plane-wave basis set was set to be 400 eV. The convergence threshold was $10^{-4}$ eV for energy and 0.02 eV/Å for force, respectively. A vacuum at least 16 Å in the z-direction was used to avoid the interaction between two periodic units. The Brillouin zone was sampled by a Monkhorst-Pack K-mesh of 3 × 3 × 1 grid for supper cell. For the calculation of containing substrate, monolayer boron on five-layer Ag (111) surface was used, where the bottom three layers were fixed. The Brillouin zone was sampled by a 1 × 1 × 1 k-point grid for supper cell and the convergence threshold for force was 0.05 eV/Å accordingly.

The HER catalytic activity of materials can be evaluated by free energy of hydrogen adsorption[28, 40], which is defined as:

$$\Delta G_H = \Delta E_H + \Delta E_{ZPE} - T\Delta S_H$$

where $\Delta E_H$ is the hydrogen adsorption energy, $\Delta E_{ZPE}$ and $\Delta S_H$ are the zero point energy and entropy differences between the adsorbed state and gas phase, respectively and $T$ is the temperature (room temperature adopted in this work). The contribution from the configurational entropy in the adsorbed state is small and can be neglected. Thus, we take the entropy of hydrogen adsorption as $\Delta S_H = \frac{1}{2} S_{H_2}$, where $S_{H_2}$ is the entropy of molecule hydrogen in the gas phase at standard condition. Besides, the hydrogen chemisorption energy is obtained via:

$$\Delta E_H = E_{(System+H)} - E_{(System)} - \frac{1}{2} E_{H_2}$$

where $E_{(system+H)}$ and $E_{(system)}$ are the energies of system with and without the adsorption of a hydrogen atom, respectively, and $E_{H_2}$ is the energy of the hydrogen molecule.

## Acknowledgements

This work is supported by the NSFC (21525311, 21373045) and NSF of Jiangsu (BK20130016) and SRFDP (20130092110029) in China. The authors thank the computational resources at the SEU and National Supercomputing Center in Tianjin.